\documentclass[ reprint, superscriptaddress, nofootinbib,
  amsmath,amssymb, aps, pra, ]{revtex4-2}

\usepackage[utf8]{inputenc}
\usepackage{textcomp}
\usepackage{amssymb, amsmath, mathtools, mathrsfs, bm, physics, amsthm}
\usepackage{bbm}
\usepackage{dcolumn}
\usepackage{dsfont}
\usepackage{MnSymbol}

\usepackage{graphicx}
\usepackage[percent]{overpic}
\usepackage{multirow}
\usepackage{verbatim}

\usepackage{xcolor}
\usepackage[normalem]{ulem}

\newtheorem{proposition}{Proposition}

\usepackage{enumitem}
\setlist[enumerate]{nosep,leftmargin=*}

\usepackage{url}
\usepackage{hyperref}
\usepackage[nameinlink,capitalize]{cleveref}
\hypersetup{
 colorlinks = true,
 linkcolor = purple,
 citecolor = purple,
 urlcolor = blue,
 linkbordercolor = white
}

\usepackage{orcidlink}

\newcommand{\eye}{\mathbbm{1}}
\newcommand{\m}[1]{\mathcal{#1}}

\usepackage{comment}

\begin{document}

\title{Generating Non-Decomposable Maps with Differentiable Semidefinite Programming}

\date{\today}

\author{Angela Rosy Morgillo\,\orcidlink{0009-0006-6142-0692}}
\email[Angela Rosy Morgillo: ]{angelarosy.morgillo01@ateneopv.it}
\affiliation{Dipartimento di Fisica, Università degli Studi di Pavia, Via Agostino Bassi 6, I-27100, Pavia, Italy}
\affiliation{INFN Sezione di Pavia, Via Agostino Bassi 6, I-27100, Pavia, Italy}

\author{Davide Poderini\,\orcidlink{0000-0003-0577-1608}}
\email[Davide Poderini: ]{davide.poderini@unipv.it}
\affiliation{Dipartimento di Fisica, Università degli Studi di Pavia, Via Agostino Bassi 6, I-27100, Pavia, Italy}

\author{Fabio Anselmi\,\orcidlink{0000-0002-0264-4761}}
\email[Fabio Anselmi: ]{fabio.anselmi@units.it}
\affiliation{Dipartimento di Matematica, Informatica e Geoscienze, Università di Trieste, via Alfonso Valerio 2, 34127 Trieste, Italy}
\affiliation{MIT, 77 Massachusetts Ave, Cambridge, MA 02139, United States of America}

\author{Fabio Benatti\,\orcidlink{0000-0002-0712-2057}}
\email[Fabio Benatti: ]{fbenatti@units.it}
\affiliation{Dipartimento di Fisica, Università di Trieste, Strada Costiera 11, 34151 Trieste, Italy}
\affiliation{INFN Sezione di Trieste, via Alfonso Valerio 2, I-34151, Trieste, Italy}

\author{Massimiliano F. Sacchi\,\orcidlink{0000-0002-8909-2196}}
\email[Massimiliano F. Sacchi: ]{massimiliano.sacchi@unipv.it}
\affiliation{CNR-Istituto di Fotonica e Nanotecnologie, Piazza Leonardo da Vinci 32, I-20133, Milano, Italy}
\affiliation{Dipartimento di Fisica, Università degli Studi di Pavia, Via Agostino Bassi 6, I-27100, Pavia, Italy}

\author{Chiara Macchiavello\,\orcidlink{0000-0002-2955-8759}}
\email[Chiara Macchiavello: ]{chiara.macchiavello@unipv.it}
\affiliation{Dipartimento di Fisica, Università degli Studi di Pavia, Via Agostino Bassi 6, I-27100, Pavia, Italy}
\affiliation{INFN Sezione di Pavia, Via Agostino Bassi 6, I-27100, Pavia, Italy}

\begin{abstract}
Positive maps that are not decomposable are a key resource in
entanglement theory because they can detect bound entangled states,
yet systematic methods for constructing them remain limited. We
introduce an optimization framework based on differentiable
semidefinite programming (SDP) for generating positive
non-decomposable maps under flexible structural constraints on their
Choi matrices. The method combines SDP-based certificates of
non-decomposability and positivity with gradient-based optimization,
enabling a systematic search over maps with different input and output
dimensions. Within this framework, we generate previously unknown
numerical examples, identify a parametrized family of maps arising
from masked Choi matrices, and construct real non-decomposable
maps. We further show that the same approach can be adapted to explore
open questions in quantum information theory, including the PPT square
conjecture and recently proposed eigenvalue bounds for 2-positive
trace-preserving maps.
\end{abstract}

\maketitle

\section{Introduction}

Positive maps that are not completely positive are fundamental tools
in quantum information theory because they provide an operational way
to detect
entanglement~\cite{bengtsson2017geometry,horodecki2009quantum,guhne2009entanglement}. A
linear map $\Phi$ is positive if it preserves positive semidefinite
operators. Although complete positivity is required for physically
realizable quantum channels, positivity alone is sufficient for
entanglement detection: if $\eye \otimes \Phi$ maps a bipartite state
$\rho$ to a non-positive operator, then $\rho$ must be entangled. The
best-known example is the Peres--Horodecki criterion based on partial
transposition~\cite{peres1996separability,HORODECKI19961}.

Among positive maps, non-decomposable ones are especially important
because they can detect entangled states with positive partial
transpose (PPT), which are invisible to decomposable
witnesses~\cite{majewski2005k}. A decomposable map can be written as
the sum of a completely positive map and a completely copositive one,
whereas a non-decomposable map cannot. Despite their central role in
entanglement theory, explicit constructions are still relatively
scarce, especially in higher dimensions, and no broadly applicable
framework for their systematic generation is currently
available~\cite{muller2016positivity,szarek2008geometry,majewski2004positive,woronowicz1975positive,choi1972,benatti2004non,stormer1982decomposable,zyczkowski2004duality,balasubramanian2020note,skowronek2009positive,ha2014global}.

Since Choi's original example~\cite{choi1975positive}, only a limited
number of explicit families of non-decomposable positive maps have
been
identified~\cite{terhal2001family,kossakowski2003class,chruscinski2009geometry,hou2010characterization,marciniak2017merging,jannesary2025class,muller2018decomposability,benatti2004quantum,majewski2017origin,bhattacharya2021generating,zwolak2013new,gulati2025positive}. Most
subsequent developments generalize or reparametrize known
constructions~\cite{mlynik2025characterization,ha2013exposedness,cho1992generalized,chruscinski2018generalizing,jafarizadeh2006generalized},
while comparatively few works address systematic generation
strategies~\cite{ende2025k,kye2013facial}. Developing flexible
numerical methods for constructing such maps is therefore of both
mathematical and physical interest.

A natural approach is to formulate the search as an optimization
problem. In this setting, semidefinite programming has become a
standard tool in quantum information because it provides efficient
convex formulations for optimization over positive semidefinite
operators~\cite{vandenberghe1996semidefinite,skrzypczyk2023semidefinite,mironowicz2024semi}. SDP-based
methods are particularly effective in problems involving quantum
channels, separability tests, and entanglement
certification~\cite{audenaert2002optimizing,doherty2004complete,doherty2014entanglement,wu2025hybrid}. More
recently, SDP techniques have also been applied to the study of
$k$-positive maps and related conjectures, including the PPT square
conjecture~\cite{ende2025k,chen2025symmetry,majewski2021ppt}.

In this work, we develop a differentiable SDP framework for generating
positive non-decomposable maps. The key idea is to treat SDP-based
tests of non-decomposability and positivity as differentiable layers
inside a gradient-based optimization loop acting directly on a
parametrized Choi matrix. This yields a systematic search procedure
that can incorporate additional requirements such as trace
preservation, reality constraints, or masked sparsity patterns.  The
framework is flexible across dimensions and can be
adapted to problem-specific objectives.

Our main contributions are as follows. First, we introduce an
optimization procedure that combines differentiable SDP layers with
standard gradient-based training to generate positive non-decomposable
maps. Then, we use this procedure to obtain new numerical examples
in dimensions between $2$ and $4$ and to identify parametrized
families arising from masked Choi matrices. Finally, we show how the
same framework can be adapted to construct real non-decomposable maps
and to explore open questions in quantum information, such as the PPT
square conjecture and an eigenvalue bound for 2-positive
trace-preserving maps. Throughout, positivity is certified through
positivity on PPT $k$-symmetrically extendable states~\cite{doherty2004complete,doherty2014entanglement}, which yields a
practical sufficient condition amenable to SDP.

The paper is organized as follows. In \cref{sec:theory}, we review the
relevant background on positive maps, decomposability, the
Choi--Jamio\l kowski isomorphism, and semidefinite programming. In
\cref{sec:methods}, we present the SDP formulations and the
differentiable optimization framework. In \cref{sec:results}, we
report the maps generated by the method, present parametrized
families obtained from masked Choi matrices, and illustrate customized loss
functions for additional tasks. Finally, in \cref{sec:conclusions}, we
summarize the main results and outline possible directions for future
work.

\section{Theory}
\label{sec:theory}

In this section, we review the concepts needed to describe and analyze positive maps in quantum information theory. We recall the notions of positivity, complete positivity, and decomposability, as well as the Choi--Jamio\l kowski isomorphism. We also briefly introduce semidefinite programming, which underlies the optimization framework developed in this work.

\subsection{Positive and non-decomposable maps}
\label{sec:theoryA}

Let $\mathcal{B}(\mathbb{C}^n)$ denote the set of $n \times n$ complex matrices. A linear map $\Phi: \mathcal{B}(\mathbb{C}^d) \to \mathcal{B}(\mathbb{C}^{d'})$ is \emph{positive} if it sends positive semidefinite operators to positive semidefinite operators, namely
\begin{equation}
\Phi(X) \succeq 0\,, \quad \forall\, X \succeq 0\, .
\label{eq:positivity}
\end{equation}
A stronger notion is \emph{$k$-positivity}: $\Phi$ is $k$-positive if its extension $\eye_k \otimes \Phi$ preserves positivity on operators acting on $\mathbb{C}^k \otimes \mathbb{C}^d$, that is,
\begin{equation}
(\eye_k \otimes \Phi)(X) \succeq 0\,, \quad \forall\, X \succeq 0\, .
\label{eq:k-positivity}
\end{equation}
Here, $\eye_k$ denotes the identity map on $\mathbb{C}^k$. A map is \emph{completely positive} (CP) if it is $k$-positive for all $k \ge 1$. A map is \emph{trace-preserving} (TP) if
\begin{equation}
\Tr[\Phi(X)] = \Tr[X]\,, \quad \forall\, X\, .
\label{eq:tp}
\end{equation}
Completely positive trace-preserving maps are quantum channels.

CP maps admit a Kraus representation~\cite{choi1975completely},
\begin{equation}
\Phi(\cdot) = \sum_l E_l\,(\cdot)\,E_l^\dagger\, ,
\label{eq:kraus}
\end{equation}
and the TP condition is equivalent to $\sum_l E_l^\dagger E_l = I_d$, where $I_d$ is the $d \times d$ identity matrix.

Positive maps display a substantially richer structure. A positive map $\Phi$ is called \emph{decomposable} if it can be written as
\begin{equation}
\Phi = \Phi_1 + \Phi_2\, ,
\label{eq:dec}
\end{equation}
where $\Phi_1$ is CP and $\Phi_2$ is completely copositive, meaning that $\Phi_2 \circ T$ is CP, with $T$ denoting transposition. Maps that do not admit such a decomposition are called \emph{non-decomposable}. These maps are of particular interest because they detect PPT entangled states that cannot be detected by decomposable maps~\cite{bengtsson2017geometry}. Moreover, for $dd' \leq 6$ all positive maps are decomposable~\cite{peres1996separability,HORODECKI19961}; non-decomposable maps therefore occur only in sufficiently high dimensions.

A linear map can be represented by its \emph{Choi matrix} $C_{\Phi}$, which establishes a one-to-one correspondence between linear maps and bipartite operators via the Choi--Jamio\l kowski isomorphism~\cite{choi1975completely}. It is defined by
\begin{equation}
\label{eq:choi}
C_{\Phi}= (\eye_d \otimes \Phi)\dyad{\psi}\,,
\end{equation}
where
\begin{equation}
\label{symmproj}
\ket{\psi}=\sum_{i=1}^{d}\ket{i} \otimes \ket{i}
\end{equation}
is the maximally entangled vector on $\mathbb{C}^d \otimes \mathbb{C}^d$.

The CJ isomorphism allows one to study linear maps through
operators. In this representation, $\Phi$ is completely positive if
and only if $C_{\Phi}$ is positive semidefinite, whereas mere
positivity is equivalent to block positivity:
\begin{equation}
 \bra{v }\otimes \bra{w} C_{\Phi} \ket{v} \otimes \ket{w} \ge 0\,, \quad \forall\, v \in \mathbb{C}^d\,,\ \forall\, w \in \mathbb{C}^{d'}\,.
 \label{eq:block_pos}
\end{equation}
Accordingly, negative eigenvalues of the Choi matrix indicate that the map cannot be completely positive, although it may still be positive.

\subsection{Differentiable semidefinite programming}
\label{sec:theoryB}

Semidefinite programming (SDP) is a standard branch of convex optimization that generalizes linear programming to the cone of positive semidefinite matrices~\cite{vandenberghe1996semidefinite,gartner2012approximation}. It is widely used in quantum information theory for numerical problems involving quantum states, channels, and separability criteria~\cite{skrzypczyk2023semidefinite}.

A standard SDP can be written as
\begin{align}
 \min_X \quad & \Tr(AX) \nonumber \\
 \text{subject to} \quad & \Tr(F_i X) = b_i\,, \quad i=1,\ldots,n \nonumber\\
 & X \succeq 0\, .
\end{align}
Here, $X$ is the positive semidefinite optimization variable, $A$ is the Hermitian objective matrix, and $(F_i,b_i)$ define potential equality constraints.

The appeal of SDP lies in its convexity and duality properties, which often permit efficient global optimization. In many applications, however, the SDP is only one component of a larger optimization problem. One may then regard the SDP as a map $f(A,F_i,b_i)$ returning the optimal value of a parametrized problem, leading to a higher-level objective of the form
\[
\min_{\theta} L\bigl(f(A(\theta),F_i(\theta),b_i(\theta))\bigr)\, .
\]
This can be computationally demanding when gradients with respect to $\theta$ are unavailable.

Recent work has shown that SDPs can be made differentiable by differentiating the optimality conditions and expressing gradients directly in terms of the input parameters~\cite{agrawal2018rewriting}. This makes it possible to incorporate SDP layers into larger optimization architectures. In the present work, we exploit this idea to explore positive non-decomposable maps by encoding positivity and non-decomposability into parametrized SDP layers embedded within a gradient-based optimization framework.

\begin{figure*}[t]
 \centering
 \includegraphics[width=\textwidth]{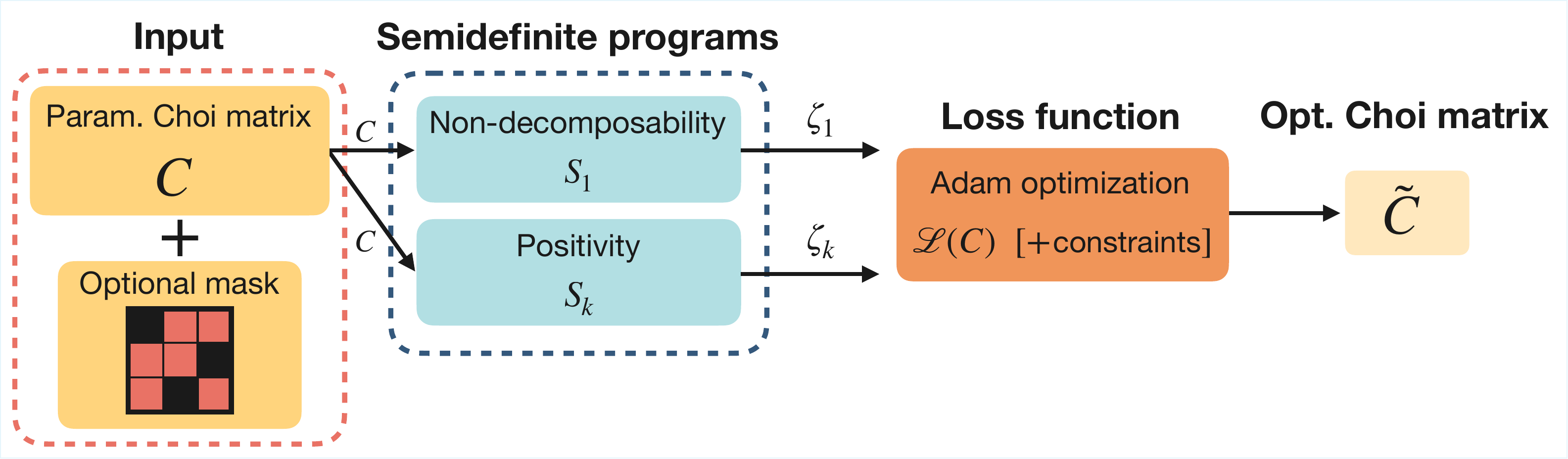}
 \caption{Schematic overview of the optimization framework. A
      parametrized Choi matrix $C$, optionally constrained by a mask,
      is processed by two SDP layers. The first evaluates the
      witness-based objective $\zeta_1$, which certifies
      non-decomposability when it becomes negative. The second
      evaluates $\zeta_k$, a positivity certificate based on PPT
      $k$-symmetric extensions. These quantities are combined in the
      loss function~\eqref{eq:loss1}, possibly together with
      additional penalty terms implementing further structural
      constraints. Minimizing this loss with Adam yields an updated
      Choi matrix $\tilde C$ whose associated map is positive and non-decomposable.}
 \label{fig:scheme}
\end{figure*}

\section{Methods}
\label{sec:methods}

In this section, we describe an SDP-based method for testing positivity of arbitrary linear maps and explain how it can be extended into a differentiable framework for generating positive non-decomposable maps.

\subsection{Semidefinite programming for positive maps}
\label{subsec:sdppos}

Starting from the Choi matrix in \cref{eq:choi}, the action of the map on a state $\rho$ is recovered as
\begin{equation}
 \Phi(\rho) = \Tr_1 \bigl[(\rho^T \otimes I_{d'}) C_{\Phi}\bigr] \, ,
\end{equation}
where $\Tr_1$ denotes the partial trace over the first subsystem, and $T$ denotes transposition with respect to the basis used in \cref{symmproj}. In what follows, we omit the subscript and write simply $C$.

Our goal is to test positivity of maps that are not assumed to be completely positive. By definition, one requires
\begin{equation}
 \Phi(\rho) = \Tr_1 \bigl[(\rho^T \otimes I_{d'}) C\bigr] \succeq 0\,, \quad \forall\, \rho\, .
\end{equation}
This can be reformulated as the optimization problem
\begin{align}
 \min_{X \succeq 0} \quad & \Tr\bigl[\Phi(\rho) X\bigr] = \Tr \bigl[(\rho^T \otimes X) C \bigr] \nonumber \\
 \text{subject to} \quad & \Tr(\rho) = 1\,,\quad \rho \succeq 0\, .
 \label{eq:positivity_opt}
\end{align}
Here, $X$ is a positive semidefinite test operator probing the spectrum of $\Phi(\rho)$. Because the objective depends bilinearly on $\rho$ and $X$, this problem is not yet an SDP.

After normalizing $X \succeq 0$, one may identify $\rho^T \otimes X$ with a product state $\tilde{\rho}=\rho_1\otimes\rho_2$. This leads to the equivalent formulation
\begin{align}
 \min_{\tilde{\rho}} \quad & \Tr(\tilde{\rho} C) \nonumber \\
 \text{subject to} \quad & \Tr(\tilde{\rho}) = 1\,,\quad \tilde{\rho} \succeq 0\,, \quad \tilde{\rho}=\rho_1 \otimes \rho_2 \, .
 \label{eq:new_positivity_opt}
\end{align}
The product-state constraint can be relaxed using the $k$-symmetric extension method~\cite{doherty2004complete,doherty2014entanglement}, which approximates the set of separable states by a convergent hierarchy of SDPs. Although the original requirement is that $\tilde{\rho}$ be a product state, it is sufficient to optimize over the larger set of separable states because the objective is linear and the minimum is attained at an extremal point, namely at a pure product state.

\paragraph*{\normalfont\bfseries $K$-symmetric extension.}

The $k$-symmetric extension criterion becomes necessary and sufficient for separability in the limit $k \to \infty$. The idea is that, given a separable state $\rho_{AB} = \sum_i p_i\, \rho_A^i \otimes \rho_B^i$, one can construct an extension to a state of $k+1$ subsystems, $\sigma_{AB_1 \ldots B_k}$, such that
\begin{equation}
 \Tr_{B_2\ldots B_k}(\sigma_{AB_1 \ldots B_k}) = \rho_{AB}\, ,
\end{equation}
where $B_1,B_2,\ldots,B_k$ are identical copies of $B$ and $\sigma_{AB_1 \ldots B_k}$ is invariant under permutations of the $B_i$ subsystems. For a separable state, one may choose
\[
\sigma_{AB_1 \ldots B_k} = \sum_i p_i\, \rho_A^i \otimes (\rho_B^i)^{\otimes k}\, ,
\]
which is manifestly symmetric. Conversely, if $\rho_{AB}$ is entangled, there exists a finite $k$ for which no such symmetric extension exists.

For fixed $k$, these conditions can be formulated as an SDP, yielding a hierarchy that increasingly approximates the set of separable states.

\paragraph*{\normalfont\bfseries SDP formulation.}

Imposing a $k$-symmetric extension allows us to minimize over a set containing all separable states, which is sufficient for the positivity test in \cref{eq:new_positivity_opt}. The resulting SDP is
\begin{align}
 \min_{\sigma \in S_k} \quad & \Tr (\tilde{\rho} C) \nonumber \\
 \text{subject to} \quad & \sigma \succeq 0\,, \quad \Tr(\sigma) = 1 \nonumber \\
 & \sigma_{AB_1 \ldots B_k} = \sigma_{AB_{\pi(1)} \ldots B_{\pi(k)}} \,, \quad \forall\, \pi \in \m P_k \nonumber \\
 & \Tr_{B_2\ldots B_k}(\sigma_{AB_1 \ldots B_k}) = \tilde{\rho}\, ,
 \label{eq:positivity_sdp}
\end{align}
where $\m P_k$ denotes the set of all permutations of $k$ indices and $S_k$ the set of $k$-symmetrically extendable states.

For finite $k$, the relaxation may include states that are not
separable. Consequently, a negative objective value does not certify
non-positivity of the map, whereas a non-negative objective value
certifies positivity. The relaxation can be strengthened by imposing
the positive partial transpose (PPT) condition on the extension
$\sigma_{AB_1\ldots B_k}$, requiring the relevant partial transposes
to remain positive semidefinite. Owing to the permutation symmetry, it
is sufficient to impose only a linear number of PPT constraints. The
resulting optimization problem is
\begin{align}
 \min_{\sigma \in S_k^{\text{PPT}}} \quad & \Tr(\tilde{\rho} C) \nonumber\\
 \text{subject to} \quad & \sigma \succeq 0\,, \quad \Tr(\sigma) = 1 \nonumber\\
 & \sigma_{AB_1 \ldots B_k} = \sigma_{AB_{\pi(1)} \ldots B_{\pi(k)}} \,, \quad \forall\, \pi \in \m{P}_k \nonumber\\
 & \sigma_{AB_1 \ldots B_k}^{T_l} \succeq 0 \, , \quad \forall\, l=1,\ldots,k \nonumber\\
 & \Tr_{B_2\ldots B_k}(\sigma_{AB_1 \ldots B_k}) = \tilde{\rho}\, ,
 \label{eq:positivity_ppt}
\end{align}
where $S_k^{\text{PPT}}$ denotes the set of PPT $k$-symmetrically extendable states and $T_l$ the partial transposition on the first $l$ subsystems $B_{i \leq l}$.

\subsection{Generating non-decomposable maps}
\label{subsec:mapgen}

We now present the optimization framework used to generate positive
non-decomposable maps. The method relies on two versions of the SDP in
\cref{eq:positivity_ppt}, corresponding to $k=1$ and $k \ge 2$.

The first SDP is
\begin{equation}
 \zeta_1(C) = \min_{\sigma \in \text{PPT}} \Tr(\sigma C)\, ,
 \label{eq:non-decsdp}
\end{equation}
and is used to detect non-decomposability. If $\zeta_1 (C)< 0$, then there exists a PPT state $\sigma$ such that $\Tr(\sigma C)<0$, implying that the associated map is non-decomposable.

The second SDP, for $k \ge 2$, is used to certify positivity on all $k$-symmetrically extendable states:
\begin{equation}
 \zeta_k(C) = \min_{\sigma \in S_k^{\text{PPT}}}\Tr(\sigma C)\,.
 \label{eq:possdp}
\end{equation}
Since the set of $k$-symmetrically extendable states contains all separable states, positivity on this relaxation implies positivity of the map. The converse, however, does not hold. The procedure therefore explores a subset of all positive maps, namely those certified by the chosen relaxation.

Both SDPs can be incorporated into a differentiable optimization loop to generate Choi matrices associated with positive non-decomposable maps. We parametrize $C$ as a $dd' \times dd'$ Hermitian matrix whose entries are trainable variables. Trace preservation can be enforced directly by requiring
\begin{equation}
 \Tr_2(C) = I_d \, ,
\label{eq:tp_condition_choi}
\end{equation}
with the explicit parametrization described in \cref{app:tp_cond}.

We then minimize the loss
\begin{equation}
 \min_C \mathcal{L}(C)\, ,
\end{equation}
where
\begin{equation}
 \mathcal{L}(C) = \text{ReLU}(\epsilon + \zeta_1(C)) + \gamma\, \text{ReLU}(-\zeta_k(C)) \, ,
 \label{eq:loss1}
\end{equation}
with hyperparameters $\epsilon > 0$ and $\gamma > 0$. Here, $\text{ReLU}(x)=\max(0,x)$. The parameter $\epsilon$ avoids the trivial solution $\zeta_1(C)=0$, while $\gamma$ controls the relative weight of the positivity term.

The first term in \cref{eq:loss1} promotes non-decomposability by
driving $\zeta_1(C)$ to negative values. The second term promotes
positivity by enforcing $\zeta_k(C)>0$. The optimization is performed
with Adam~\cite{kingma2014adam}, and the SDPs are implemented as
differentiable layers using CVXPY and
PyTorch~\cite{NEURIPS2019_9ce3c52f,agrawal2018rewriting,diamond2016cvxpy}. The
framework can be customized further by adding new loss terms or by
imposing masks on the Choi matrix, as discussed below. A schematic
summary is shown in \cref{fig:scheme}.

Although the SDP-based procedure is designed for non-decomposable
maps, decomposable positive maps can be generated more directly and do
not require SDP. For completeness, we give a simple construction in
\cref{app:dec_algo}.

\begin{table}[thb]
 \centering
 \begin{tabular}{cccrrr}
 \hline
 $d$ & $d'$ & $k$ & ASE & Success rate (\%) \\
 \hline
 2 & 4 & 2 & 434.7 $\pm$ 35.8 & 27 \\
 4 & 2 & 2 & 394.0 $\pm$ 36.6 & 23 \\
 3 & 3 & 2 & 217.8 $\pm$ \ 5.4 & 98.4 \\
 4 & 4 & 2 & 280.5 $\pm$ 11.3 & 100 \\
 \hline
 \end{tabular}
 \caption{Performance of the algorithm in generating positive
   non-decomposable maps without imposing trace preservation. For each
   choice of input dimension $d$, output dimension $d'$, and order
   $k$, we report the average success epoch (ASE), defined as the mean
   epoch at which the loss function in \cref{eq:loss1} vanishes,
   together with its standard deviation, and the success rate, defined
   as the fraction of successful runs. Unless otherwise stated, the
   algorithm was run 200 times for at most 2000 epochs with
   hyperparameters $\varepsilon = 0.05$, $\gamma = 2$, and learning
   rate $\eta = 0.01$.}
 \label{tab:res_summary}
\end{table}

\section{Results}
\label{sec:results}

We now present the results obtained with the optimization framework introduced above.

We first consider the generation of positive non-decomposable maps
without additional structural constraints. This is achieved by
minimizing the loss in \cref{eq:loss1}. Whenever the optimization
converges to a Choi matrix $C^*$ such that $\mathcal{L}(C^*)=0$, the
associated map is certified to be positive and
non-decomposable. Repeating the procedure with different random
initializations and hyperparameter values yields multiple distinct
examples.

The performance of the algorithm for input and output dimensions
between $2$ and $4$, with order $k=2$, is summarized in
\cref{tab:res_summary}. In the symmetric cases $d=d'$, the
optimization reliably identifies positive non-decomposable maps in
essentially every run. In contrast, the success rate is lower for
asymmetric dimensions. The origin of this behavior is not yet
clear. One possible explanation is that the $k$-symmetric extension is
imposed on the subsystem with smaller dimension, leading to a faster
but less accurate relaxation. In principle, increasing $k$ could
improve the approximation, but this is currently limited by
computational cost.

For $d=d'=3$, we performed $130$ runs, whereas for $d=d'=4$ we
performed $30$ runs, since the success rate was consistently $100\%$
and a smaller sample sufficed to verify the stability of the
result. By contrast, for the asymmetric cases we used $200$ runs to
obtain more reliable estimates of both the success rate and the
standard deviation of the ASE.

Both the success rate and the convergence time depend on the choice of
hyperparameters $\epsilon$ and $\gamma$. For the case $d=d'=3$, a more
detailed numerical analysis of this dependence is reported in
\cref{app:convergence_analysis}. The randomness of the procedure
originates from the initial seed defining the starting Choi matrix
$C_0$, which is taken to be a generic Hermitian matrix.

\subsection{Masked Choi matrices and a new family of maps}
\label{subsec:bound_mask}

The optimization problem can be simplified by imposing additional
structure on the Choi matrix, thereby reducing the number of free
parameters. To this end, we introduce heuristic masks that force
selected entries of the Choi matrix to vanish. Although these masks
are not derived from first principles, they substantially reduce the
complexity of the search space and can reveal analytically tractable
structures.

Using this strategy, we identified a family of maps with Choi matrices of the form
\begin{equation}
 C=\left(\begin{array}{ccc|ccc|ccc}
 a & \cdot & \cdot & \cdot & \cdot & \cdot & \cdot & \cdot & z\\
 \cdot & b & \cdot & w & \cdot & \cdot & \cdot & \cdot & \cdot\\
 \cdot & \cdot & c & \cdot & \cdot & \cdot & \cdot & \cdot & \cdot\\ \hline
 \cdot & w^* & \cdot & c & \cdot & \cdot & \cdot & \cdot & \cdot\\
 \cdot & \cdot & \cdot & \cdot & a & \cdot & \cdot & \cdot & \cdot\\
 \cdot & \cdot & \cdot & \cdot & \cdot & b & \cdot & \cdot & \cdot\\ \hline
 \cdot & \cdot & \cdot & \cdot & \cdot & \cdot & b & \cdot & \cdot\\
 \cdot & \cdot & \cdot & \cdot & \cdot & \cdot & \cdot & c & \cdot\\
 z^* & \cdot & \cdot & \cdot & \cdot & \cdot & \cdot & \cdot & a
\end{array}
\right)\, ,
\label{eq:param0}
\end{equation}
where $a,b,c$ are non-negative real parameters and $w,z$ are complex numbers. Dots denote zero entries for readability.

The corresponding action on a density matrix $\rho$ is
\begin{widetext}
\begin{equation}
 \Phi(\rho) = \begin{pmatrix}
a\rho_{11}+c\rho_{22}+b\rho_{33}
& w^*\,\rho_{21}
& z\,\rho_{13} \\[4pt]
w\,\rho_{12} & b\rho_{11}+a\rho_{22}+c\rho_{33}
& 0 \\[4pt]
z^*\,\rho_{31} & 0 &
c\rho_{11}+b\rho_{22}+a\rho_{33}
\end{pmatrix} \, .
 \label{eq:map}
\end{equation}
\end{widetext}

A complete analytical characterization of this family, including
conditions for positivity, non-decomposability, and violation of the
bound in \cref{eq:bound}, is deferred to a separate
work~\cite{new_family_sacchi}.

\begin{figure}[t]
 \centering
 \includegraphics[width=0.9\linewidth]{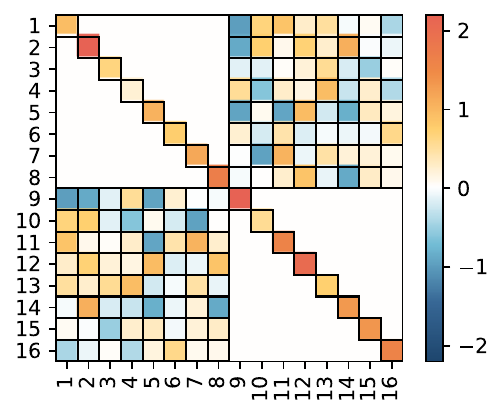}
 \caption{Example of a Choi matrix associated with a positive non-decomposable map on $\mathbb{C}^4 \otimes \mathbb{C}^4$.}
 \label{fig:example4x4}
\end{figure}

We also provide explicit examples for the case $d=d'=4$, where
comparatively few constructions are available in the
literature~\cite{robertson1985positive,hall2006new,breuer2006optimal}. The
resulting Choi matrices are shown in \cref{fig:example4x4}, together
with the mask used during the optimization.

\subsection{Customized loss function and eigenvalue bound violation}
\label{subsec:bound}

The flexibility of the framework allows additional constraints to be
incorporated directly into the loss function. As an illustration,
consider the recently proposed bound for 2-positive trace-preserving
maps~\cite{boundoneigs2025}.

\begin{proposition}[\textbf{Bound for 2-positive, trace-preserving maps}]
Let $\Phi : \mathcal{B}(\mathbb{C}^d) \to \mathcal{B}(\mathbb{C}^{d})$ be a 2-positive, trace-preserving map. Then
\begin{equation}
 \Tr \Phi \leq d\, \min \mathrm{Re}[\sigma(\Phi)] + d^2 - d \, ,
 \label{eq:bound}
\end{equation}
where $\mathrm{Re}[\sigma(\Phi)]$ denotes the real part of the eigenvalues of $\Phi$.
\end{proposition}

It was shown that this inequality is violated by the transposition map
for $d=2$, whereas for $d=3$ no violating examples are currently
known. Ref.~\cite{boundoneigs2025} conjectured that such violations
might occur for non-decomposable maps. The present framework makes it
possible to search directly for examples.

To this end, we augment the loss in \cref{eq:loss1} by adding a term that enforces violation of \cref{eq:bound}:
\begin{align}
\mathcal{L} (C) &= \text{ReLU}(\epsilon + \zeta_1(C)) + \gamma \, \text{ReLU}(\delta-\zeta_k(C)) \nonumber \\
 &\qquad + \omega\,\text{ReLU}(\nu + \xi(C)) \, ,
 \label{eq:lossbound}
\end{align}
where
\[
\xi(C) = -\Tr \Phi + d\min \text{Re}[\sigma(\Phi)] + d^2 - d\, ,
\]
and $\omega,\nu >0$. The additional hyperparameter $\delta > 0$ is
introduced to facilitate convergence toward positive maps. This loss
can produce both non-decomposable and decomposable examples: when
$\zeta_1<0$ the resulting map is non-decomposable, while cases with
$\zeta_1>0$ provide decomposable maps that nevertheless violate the
bound.

An explicit example of such maps for $d=3$ is given by
\cref{eq:param0}, with $c=1-a-b$ to ensure trace preservation.  In
particular, the choice $a=1$, $b=c=0$, and $w=z=\sqrt{2}/2$ yields an
example of a decomposable map~\cite{new_family_sacchi}.

\begin{figure*}[htb]
 \centering
 \begin{minipage}{0.3\textwidth}
 \begin{overpic}[width=\textwidth]{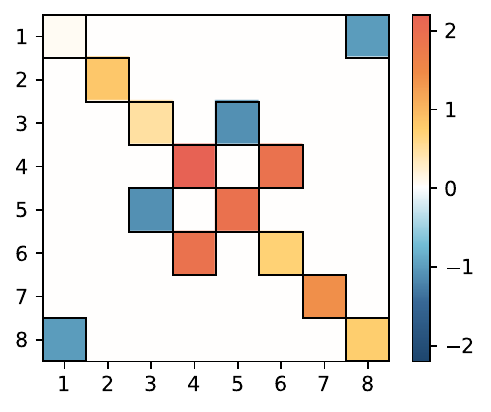}
 \put(50,-1){\makebox(0,0){\textbf{(a)}}}
 \end{overpic} \\[2mm]
 \begin{overpic}[width=\textwidth]{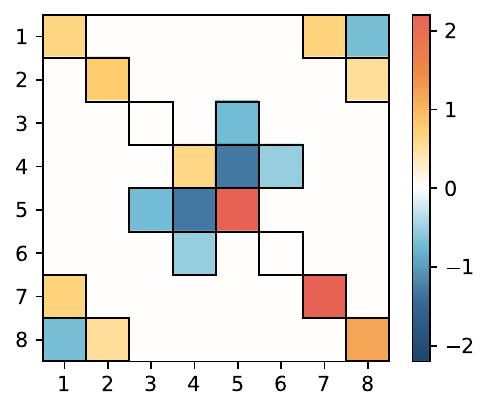}
 \put(50,-1){\makebox(0,0){\textbf{(d)}}}
 \end{overpic}
 \end{minipage}\hspace{0.02\textwidth}
 \begin{minipage}{0.3\textwidth}
 \begin{overpic}[width=\textwidth]{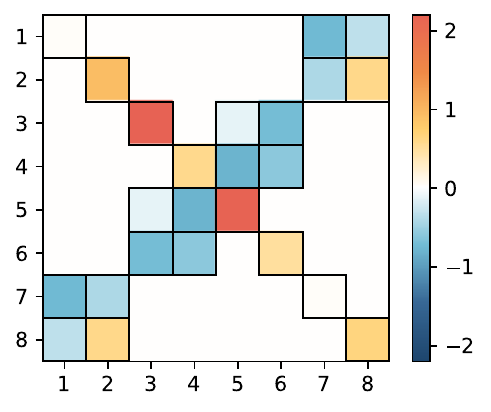}
 \put(50,-1){\makebox(0,0){\textbf{(b)}}}
 \end{overpic} \\[2mm]
 \begin{overpic}[width=\textwidth]{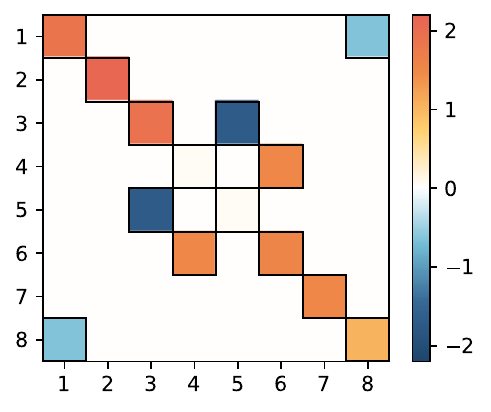}
 \put(50,-1){\makebox(0,0){\textbf{(e)}}}
 \end{overpic}
 \end{minipage}\hspace{0.02\textwidth}
 \begin{minipage}{0.3\textwidth}
 \begin{overpic}[width=\textwidth]{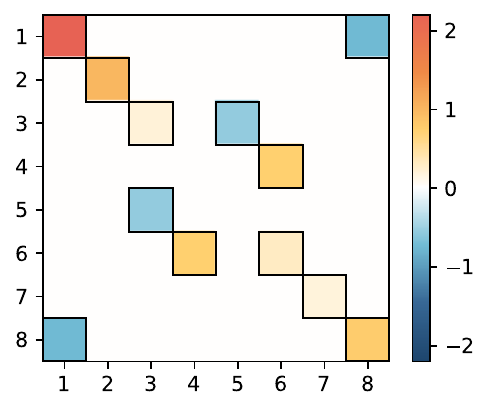}
 \put(50,-1){\makebox(0,0){\textbf{(c)}}}
 \end{overpic} \\[2mm]
 \begin{overpic}[width=\textwidth]{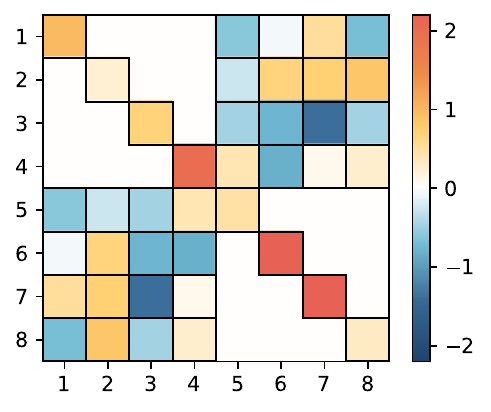}
 \put(50,-1){\makebox(0,0){\textbf{(f)}}}
 \end{overpic}
 \end{minipage}
 \caption{Examples of real Choi matrices corresponding to non-decomposable maps on $\mathbb{C}^2 \otimes \mathbb{C}^4$, obtained using different masking patterns in the optimization procedure. Nonzero entries indicate the locations selected by the masks.}
 \label{fig:real_masks}
\end{figure*}

\subsection{Construction of real non-decomposable maps}
\label{subsec:real_maps}

We now present an explicit construction relevant to the mutual
exclusivity condition discussed in
Ref.~\cite{skowronek2009positive}. The result can be stated as
follows.

\begin{proposition}[\textbf{Real non-decomposable matrices}]
\label{prop:real_mats}
Let $m \in \mathbb{N}$. Then either every block-positive operator on $\mathbb{C}^2 \otimes \mathbb{C}^m$ with real matrix elements is decomposable, or there exists a block-positive operator $A \in \mathbb{C}^2 \otimes \mathbb{C}^m$ with real entries $A_{abcd}$ that is non-decomposable and can be written as a sum of squares of bilinear forms.
\end{proposition}

For $m=1,2,3$, the statement is trivial because all positive maps are decomposable in these dimensions~\cite{peres1996separability,HORODECKI19961}. The situation becomes nontrivial for $m \geq 4$, where the existence of non-decomposable positive maps with real Choi matrices is no longer automatic.

Using the optimization procedure based on the loss in \cref{eq:loss1}, together with the constraint that the Choi matrix be real-valued, we explicitly construct such maps for $m=4$ and $m=5$. These constructions extend directly to larger values of $m$. Several examples for $m=4$ obtained with different masks are shown in \cref{fig:real_masks}. The existence of real non-decomposable positive maps was previously established by other methods~\cite{tang1986positive,ha2016construction}; here we recover such examples within a unified optimization framework.

\subsection{PPT square conjecture}
\label{subsec:ppt}

We conclude the results section by applying the framework to the PPT square conjecture~\cite{majewski2021ppt}.

\begin{proposition}[\textbf{PPT square conjecture}]
\label{prop:pps_conjecture}
Let $T_1$ and $T_2$ be positive maps, with $T_2$ being PPT (that is, both completely positive and completely copositive). Then the composition
$T_1 \circ T_2$
is necessarily decomposable.
\end{proposition}

To test this conjecture numerically, we design an optimization
strategy adapted to the present SDP framework. The Choi matrix of the
composed map $T_1 \circ T_2$ is evaluated using the SDP in
\cref{eq:non-decsdp}, which tests for non-decomposability. At the same
time, the map $T_2$ is constructed to be PPT by design. Specifically,
we choose a Hermitian matrix $A$ and define $T_2$ through a Choi
matrix of the form $C_{T_2}=AA^\dagger$, which guarantees complete
positivity. We then enforce positivity of the partial transpose by
including in the loss a term that drives all eigenvalues of the
partial transpose of $C_{T_2}$ to non-negative values. The map $T_1$
is generated using the SDP in \cref{eq:possdp}.

We focus on the case $T_1 : \mathcal{B}(\mathbb{C}^2) \to
\mathcal{B}(\mathbb{C}^4)$ and $T_2 : \mathcal{B}(\mathbb{C}^4) \to
\mathcal{B}(\mathbb{C}^2)$, since the case $T_1,T_2 :
\mathcal{B}(\mathbb{C}^3) \to \mathcal{B}(\mathbb{C}^3)$ has already
been settled~\cite{christandl2019composed}. In all cases explored, the
optimization did not reveal any violation of the conjecture: the
composed maps $T_1 \circ T_2$ were always found to be decomposable. Although this does
not constitute a proof, it provides additional numerical evidence
supporting the conjecture in the tested dimensional regime.

\section{Conclusion}
\label{sec:conclusions}

We have introduced a numerical framework for generating positive
non-decomposable maps based on differentiable semidefinite
programming. By embedding SDP-based certificates of positivity and
non-decomposability into a gradient-based optimization loop, the
method enables a systematic exploration of classes of maps that are
otherwise difficult to construct analytically.

A central strength of the framework is its flexibility. Additional
structural information can be incorporated through tailored loss
functions or through masks imposed on the Choi matrix, allowing the
search to be adapted to specific mathematical or physical
questions. We illustrated this versatility by generating new maps,
constructing examples with real Choi matrices, and probing open
problems such as eigenvalue bounds for 2-positive trace-preserving
maps and the PPT square conjecture.

We emphasize that positivity is certified through positivity on
$k$-symmetrically extendable states, which is a sufficient but
generally stronger condition than positivity itself. As a result, the
procedure may exclude valid positive maps that lie outside the chosen
relaxation. Improving the quality of this relaxation, extending the
method to larger values of $k$, and applying the framework to other
open questions in quantum information theory are natural directions
for future work.

\section*{Acknowledgements}
A.R.M. acknowledges support from the PNRR MUR Project PE0000023-NQSTI. This work has been sponsored by PRIN MUR Project 2022SW3RPY.

\section*{Code Availability}
The code underlying the results presented in this work is openly available on GitLab~\cite{podemorrepo}.

\appendix

\section{Decomposable maps generation}
\label{app:dec_algo}

In this appendix, we describe a simple algorithm for generating positive trace-preserving decomposable maps. Any such map $\Phi$ can be written as
\begin{equation}
 \Phi = p \,\Gamma_1 + (1-p)\,\Gamma_2 \circ T \, ,
 \label{eq:decv2}
\end{equation}
where $\Gamma_1$ and $\Gamma_2$ are completely positive trace-preserving (CPTP) maps and $p \in [0,1]$. Constructing a decomposable map therefore reduces to constructing two CPTP maps.

Each map $\Gamma_i$ is implemented through a trainable unitary
\[
U_i=\exp(A_i-A_i^\dagger)
\]
acting on the system together with an ancilla. Preparing the ancilla in the state $\ket{0}_A$ induces the channel
\begin{equation}
 \Gamma_i(\rho)
= \Tr_A\!\left[ U_i(\rho\otimes\dyad{0}_A)U_i^\dagger \right] \, .
\end{equation}
Evaluating the partial trace in an orthonormal basis $\{\ket{j}_A\}$ of the ancilla yields Kraus operators
\begin{equation}
K_j^{(i)} = (\eye \otimes \bra{j}_A)\, U_i\, (\eye \otimes \ket{0}_A)\,,
\end{equation}
so that
\[
\Gamma_i(\rho)=\sum_j K_j^{(i)} \rho K_j^{(i)\dagger}\,.
\]

Starting from such a decomposition, we optimize the parameters so that the Choi matrix $C$ of the resulting map has at least one negative eigenvalue, thereby ensuring that the map is not completely positive. If $\{\lambda_k\}$ are the eigenvalues of $C$, we define the loss
\begin{equation}
\mathcal{L}_{\rm P} = \operatorname{ReLU}\!\Big( \min_k \lambda_k \Big)\,.
\label{eq:losspos}
\end{equation}
The parameters of the Kraus operators, together with the mixing weight
$p \in [0,1]$, are then updated using Adam until
convergence. Repeating the optimization with different random
initializations yields an ensemble of decomposable maps for further
analysis.

\section{TP condition}
\label{app:tp_cond}
In this appendix, we explain how the trace-preserving (TP) constraint
is enforced in the optimization procedure. We first considered
imposing trace preservation through a soft penalty term added to the
loss,
\[
\mathcal{L}_{\text{TP}} = \left\| \Tr_2(C) - I_d \right\|_2\, .
\]
In practice, however, this choice led to unstable optimization and
poor convergence. We therefore impose the TP condition by
construction.

To this end, we introduce a real-valued tensor of trainable parameters
$X_{ijk\ell}$, with indices $i,k = 1,\dots,d$ and $j,\ell =
1,\dots,d'$, and use it to parametrize the Choi matrix elements
\[
C_{ijk\ell} \equiv \langle i | \otimes \langle j | C | k \rangle
\otimes | \ell \rangle
\]
as
\begin{align}
\operatorname{Re}[C_{ijk\ell}] &= \frac{1}{2}\bigl(X_{ijk\ell} + X_{k\ell ij}\bigr) \, , \\
\operatorname{Im}[C_{ijk\ell}] &= \frac{1}{2}\bigl(X_{ijk\ell} - X_{k\ell ij}\bigr) \, ,
\label{eq:parametrization}
\end{align}
which guarantees Hermiticity, namely $C_{ijk\ell} = C^*_{k\ell ij}$.

Equivalently, $C$ can be viewed as a block matrix $C=[C_{ik}]$, where each block $C_{ik} \!\in\! \mathcal{B}(\mathbb{C}^{d'})$ has entries $[C_{ik}]_{j\ell}=C_{ijk\ell}$. In this representation, the trace-preserving condition \cref{eq:tp_condition_choi} becomes
\begin{equation}
\operatorname{Tr}[C_{ik}] = \sum_{j=1}^{d'} C_{ijkj} = \delta_{ik} \, , \quad \forall\, i,k\, ,
\label{eq:tpcondC}
\end{equation}
which is linear in the diagonal entries $X_{ijkj}$ of the parameter tensor.

\begin{figure}[t]
 \centering
 \includegraphics[width=0.9\linewidth]{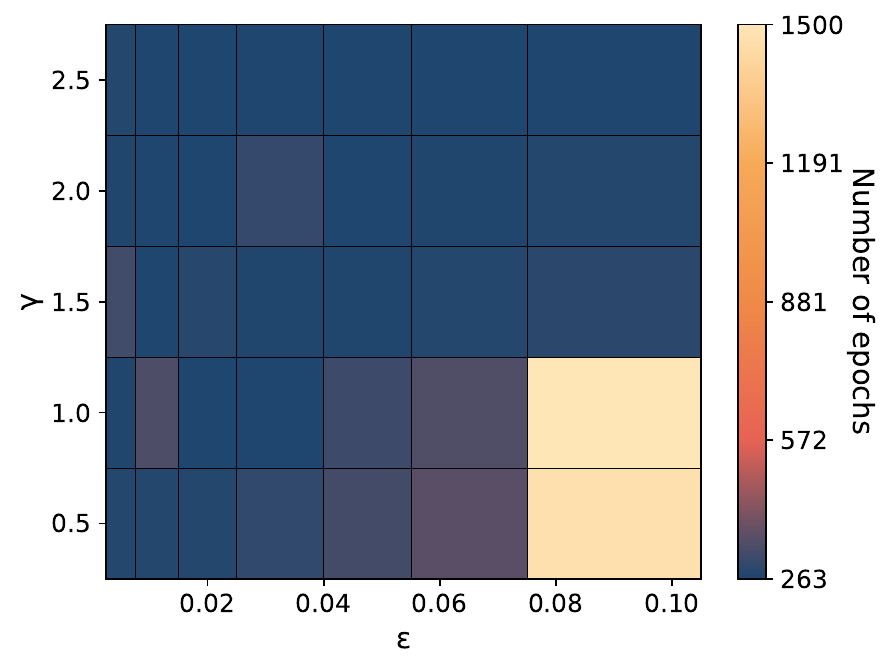}
 \caption{Number of epochs required to reach convergence, defined as the point at which the loss function in \cref{eq:loss1} becomes zero, as a function of the hyperparameters $\gamma$ and $\epsilon$.}
 \label{fig:convergence}
\end{figure}

To satisfy these constraints explicitly, we treat $d'-1$ diagonal
elements in each block $[X_{ik}]$ as free parameters and determine the
remaining one so that
\begin{equation}
\sum_{j=1}^{d'} X_{ijkj} = \delta_{ik} \, , \quad \forall\, i,k\, .
\end{equation}

\begin{figure*}[thb]
 \centering
 \includegraphics[width=\linewidth]{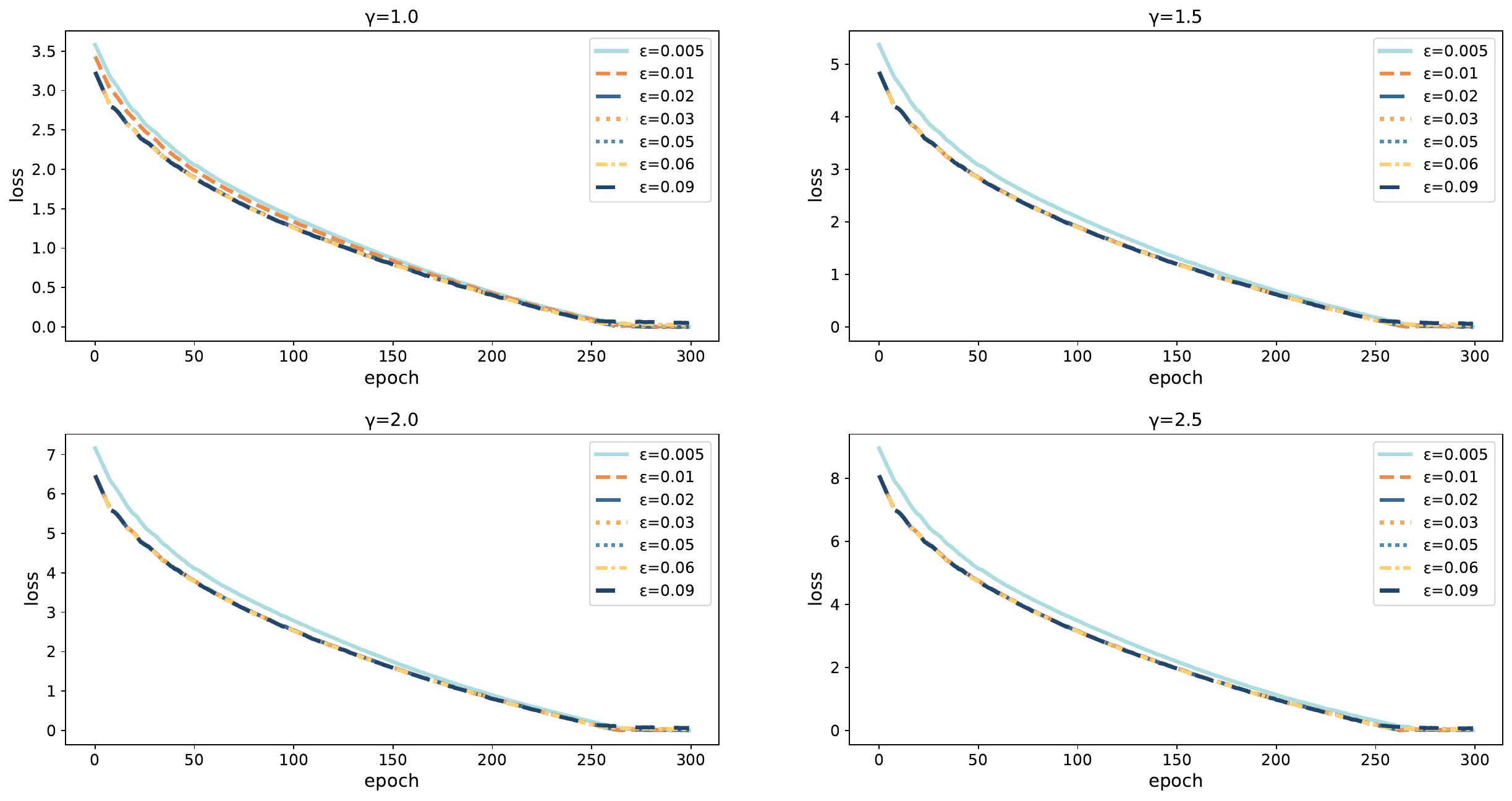}
 \caption{Evolution of the loss in \cref{eq:loss1} over the training epochs for different values of $\epsilon$ at fixed $\gamma$.}
 \label{fig:loss_eps}
 \centering
 \includegraphics[width=\linewidth]{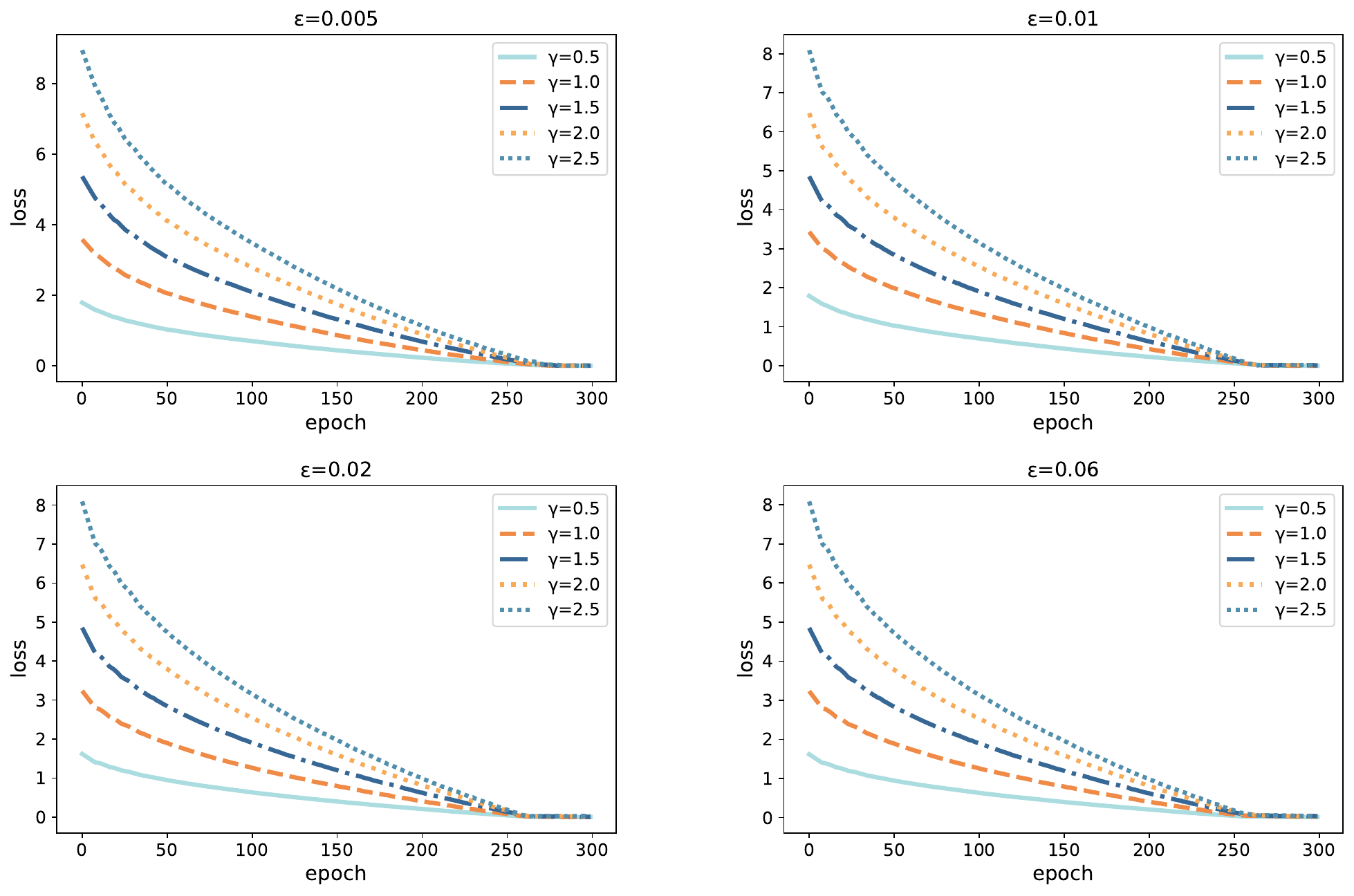}
 \caption{Evolution of the loss in \cref{eq:loss1} over the training epochs for different values of $\gamma$ at fixed $\epsilon$.}
 \label{fig:loss_gamma}
\end{figure*}

\section{Convergence analysis}
\label{app:convergence_analysis}

In this appendix, we study the convergence properties of the algorithm
as a function of the hyperparameters, focusing on the case $d=d'=3$
and $k=2$.

\cref{fig:convergence} shows the number of epochs required for
convergence, defined as the point at which the loss in \cref{eq:loss1}
reaches zero, as a function of $\gamma$ and $\epsilon$. For most
parameter choices, convergence is achieved in fewer than $400$
epochs. There are, however, combinations with small $\gamma$ and large
$\epsilon$ for which convergence is not observed. This is consistent
with the interpretation of the loss: increasing $\epsilon$ imposes a
stronger requirement on non-decomposability, while decreasing $\gamma$
weakens the relative pressure toward positivity, making it more
difficult to identify a positive non-decomposable map.

We also examine how $\epsilon$ and $\gamma$ affect the loss trajectory
during training. \cref{fig:loss_eps,fig:loss_gamma}
show the evolution of the loss for different values of $\epsilon$ and
$\gamma$, respectively. In the parameter range considered here, the
convergence behavior appears more sensitive to variations in $\gamma$
than to variations in $\epsilon$.

\begin{figure*}[htb]
 \centering
 \includegraphics[width=\linewidth]{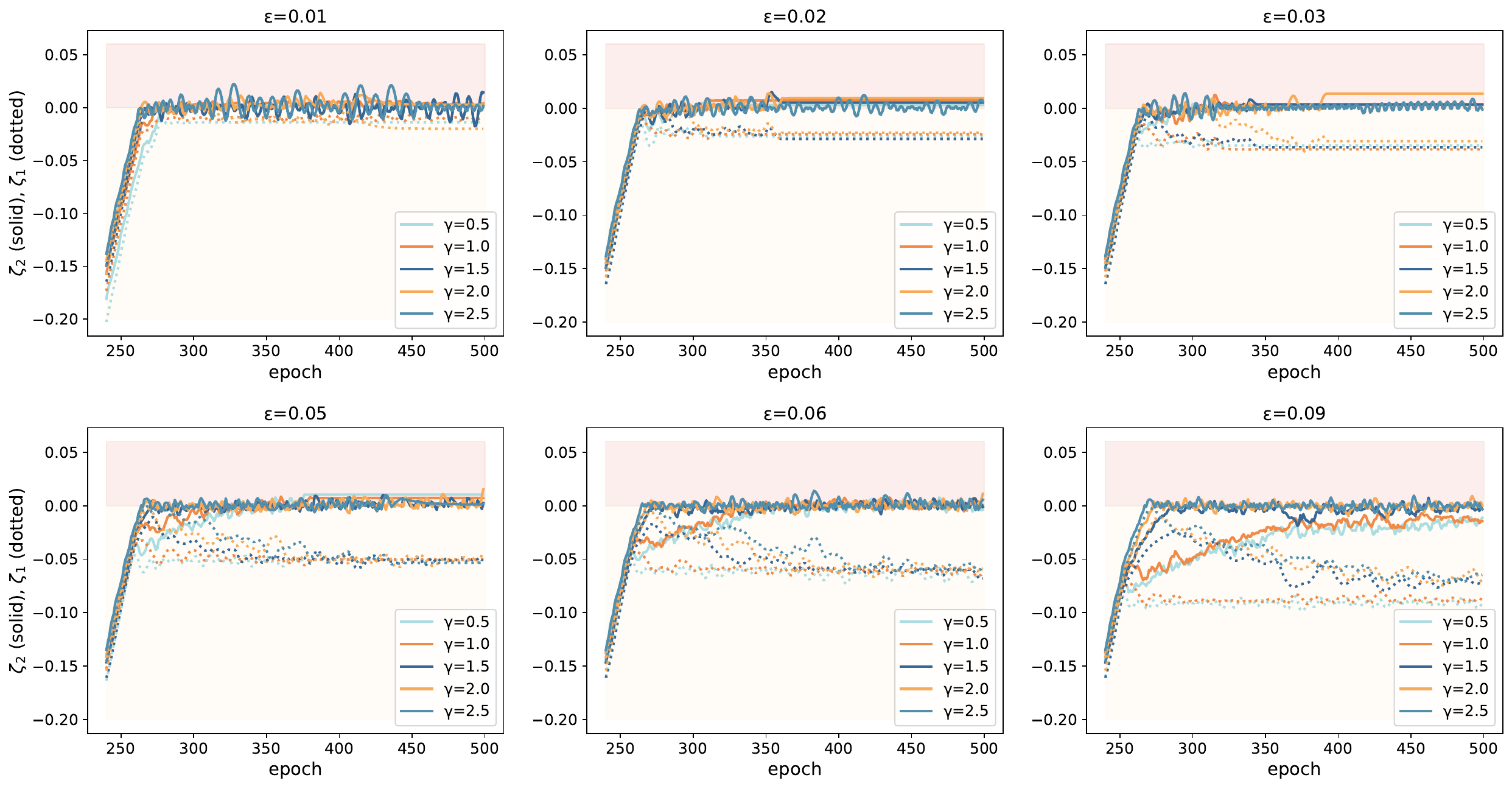}
 \caption{Evolution of the SDP objectives $\zeta_1$ and $\zeta_2$ in \cref{eq:loss1} during training.}
 \label{fig:loss_zeta}
\end{figure*}

Finally, we analyze the evolution of the SDP objectives $\zeta_1$ and
$\zeta_2$ defined in \cref{eq:non-decsdp,eq:possdp}. This makes it
possible to track how the optimization balances the two
constraints. As shown in \cref{fig:loss_zeta}, both objectives
typically start from strongly negative values. During training, around
epochs $250$--$300$, $\zeta_2$ becomes positive, indicating that the
optimization has successfully enforced positivity on the chosen
relaxation, while $\zeta_1$ remains negative, thereby certifying
non-decomposability.

\bibliography{bibliography4.bib}

\end{document}